# Indicators of Structural Change in the Dynamics of Science: Entropy Statistics of the *SCI Journal Citation Reports*

*Scientometrics* 53(1) (2002) 131-159 (*forthcoming*)

Loet Leydesdorff
Science & Technology Dynamics,
Amsterdam School of Communications Research (ASCoR),
Kloveniersburgwal 48, 1012 CX  Amsterdam, The Netherlands
loet@leydesdorff.net; http://www.leydesdorff.net/

**Abstract**

Can change in citation patterns among journals be used as an indicator of structural change in the organization of the sciences? Aggregated journal-journal citations for 1999 are compared with similar data in the *Journal Citation Reports* 1998 of the *Science Citation Index.* In addition to indicating local change, probabilistic entropy measures enable us to analyze changes in distributions at different levels of aggregation. The results of various statistics are discussed and compared by elaborating the journal-journal mappings. The relevance of this indicator for science and technology policies is further specified.

## 1. Introduction

Scientific novelty and science-based innovations can be considered as emerging phenomena. In some cases the new developments will branch out of science and technology into fields of application, but one can also expect a feedback into the further development of the relevant sciences and science-based technologies (Rosenberg, 1982; Narin & Noma, 1985). The dynamics of science-based innovation systems are then reflected in the scientific literature. For example, the discovery of superconductivity at relatively high temperatures (1987) and the oncogene (1987) occasioned the production of new journals with specific and/or changing cluster structures (Vlachý, 1988; Leydesdorff *et al.*, 1994).

Previous attempts to use the statistical properties of new journals (e.g., aggregated citation distributions) as an indicator of change over the file failed because journals are added to the ISI-database for a variety of reasons (Garfield, 1990). Scientometric networks of links can be analyzed in terms of their structural properties ('eigenvectors') and/or in terms of hierarchical relations among the journals which are located at the nodes of the network. The structural (or positional) analysis of the network is conceptually different from the use of scientometric indicators for hierarchical rank ordering of journals (Burt, 1982; Leydesdorff, 1995).

In the case of ranking one attributes the indicators to a hierarchy among the agents. Agents can be journals (e.g., Doreian, 1986), individual scientists or institutions (e.g., Irvine & Martine, 1983; Moed *et al.,* 1985). In the case of positioning the communicative operation at the link (e.g., citation) is used as the unit of analysis for studying properties of the network level (e.g., Tijssen *et al.*, 1987). The network of relations among actors generates and reproduces the communication structure as a relevant *environment* for the performing actors (Luhmann, 1984; Leydesdorff, 2001).

Scientists *compete* as agents for publication space and recognition in scientific journals at a next-order level. Decisions on whether or not to publish a given paper are made at this level, namely, by editors using the peer review system. From this perspective, editors of scientific journals have also been studied as the 'gate-keepers' of science (Zsindely *et al.*, 1982). Perhaps even to a larger extent than in the case of co-citation and co-word mappings which reflect partially the intentional constructions of the authors, the emerging network of aggregated relations among scientific journal articles are increasingly beyond the control of individual agency (Price, 1965; Leydesdorff, 1987).

Specialties can be expected to use specific (jargonistic) words and to cite in specialist domains. Citations are known to be even more highly codified than title words (Leydesdorff, 1989). With further codification and consequent delineation, developments at the structural level of aggregated journal-journal citations can be expected to exceed increasingly the control of intentional action by participants. Accordingly and at the methodological level, the graph analytical approach focuses on the hierarchy of relations in the historical construction of the system, whereas the factor analysis decomposes the latent network structures which have emerged from the construction, but from a hindsight perspective. How is action selected by the network structures that emerge from the aggregated distributions of similar actions? How can these latent structures and change at this level be indicated?

In other words, the factor analysis of aggregated journal-journal citations relations provides us with a relatively independent baseline for the measurement of scientific developments (cf. Studer and Chubin, 1980, at p. 269). The journal networks contain a relatively independent dynamics. However, when scientific fields become 'hot' because of discoveries, new scientific theories and/or technological breakthroughs, both authors and publishing houses may try to jump on the bandwagon (Leydesdorff *et al.*, 1994). Analogously, scientific specialties may lose relative relevance over time and consequently exhibit decline and/or disintegration.

In this study, I raise the question of whether, and if so how, the changes in citation patterns among scientific journals can be used as indicators of structural change in the organization of the sciences in terms of these networked journal relations. Particularly, the focus will be on the question of whether we can measure the 'heat' of a development by using probabilistic entropy as a measure of change between years. This question is particularly relevant for science and technology policy issues *at the aggregated and strategic level*. Therefore, I will focus on the present situation using the latest available data from the *Science Citation Index*. Where are new journal clusters emerging? Do these changes indicate scientific and technological breakthroughs and can they perhaps be used as early warning indicators?

## 2. Journal mapping

The *Journal Citation Reports* of the *Science Citation Index* enable us to operationalize the measurement of the relations among journals. This data is available in electronic format and on a yearly basis since 1994. The turnover of journals between the years is specified by the ISI in an additional file. In the meantime, this data provides us with a sufficient number of years for addressing questions of structural change systematically.

Leydesdorff *et al.* (1994) have used similar data for the historical monitoring and evaluation of priority areas. Journal maps were composed for different years and the comparison among these 'snapshots' were evaluated with reference to the dynamics under study (Leydesdorff & Gauthier, 1996; Van den Besselaar & Leydesdorff, 1996; Leydesdorff & Van den Besselaar, 1997). Methodologically, however, such an approach can be considered as a *comparative static analysis*. The indicator itself does not map change, but the different mappings are juxtaposed and then 'subtracted.' However, one is not able to distinguish whether the difference indicates structural change or variation.

In this context, Leydesdorff & Cozzens (1993) have proposed 'central tendency journals' as yardsticks for measuring structural change. 'Central tendency journals' are defined as seed journals that exhibit the highest correlation with the eigenvector that represents their cluster at the network level. Central tendency journals exhibit more stability than journals that are less central to the cluster. One conclusion of these previous studies has been that new developments can be indicated by journals which exhibit this structural property in the *cited* dimension. 'Citing' can be considered as the action parameter, while 'cited' reveals codification. Changes in the *cited structure* (operationalized in terms of the 'central tendency journals') indicate changes in the perception of and recognition by citing authors (Small, 1978; Leydesdorff, 1998).

A drawback of the factor analytic approach is that one cannot generalize over the file for computational reasons. Each analysis of the factor structure requires a new relational delineation of the citation environment of the seed journal. One then needs theoretical information to know 'where to look'. A graph analytical approach, on the other hand, can inform us about the hierarchical stratification in the entire database, but not specifically about the eigenstructure of the network. Specialist journals—which are not necessarily on top in the hierarchy—sometimes indicate intellectually important dimensions. The intellectual organization of the sciences is more dynamic than its reflection in prevailing organizational formats (Whitley, 1984; Luhmann, 1990).

### 3. The evolutionary perspective

Where the journal-journal citation structures become 'hot', an increase in the probabilistic entropy can be expected locally. Is it possible to specify a method for systematically spotting this 'heat' in the development of the database (cf. Kostoff, 1997)? While journal-journal mappings hitherto have used the geometrical metaphor of multi-variate spaces for *comparative static* analysis (Doreian and Ferrero, 1985; Tijssen, 1987 and 1992; Van den Besselaar, 2000), entropy statistics or 'information calculus' (Bar-Hillel, 1955) enables us to focus on change as a *dynamic and evolutionary* operation.

Since the entropy measure is composed of straightforward summations, this measure can be developed over the file using a stepwise procedure. In other words, the analyst does not have to make an initial decision about the focus, but can sort individual journals in terms of how much they contribute to change both in the cited and the citing dimension, as well as in terms of the cited/citing interaction.

Following Shannon (1948), Theil (1972) defined the expected information content *I* of a message that an *a priori* distribution $\Sigma p_i$ has turned into an *a posteriori* distribution $\Sigma q_i$, as follows:

$$I = \Sigma_i \, q_i \, {}^2\!\log (q_i / p_i)$$

When the two-base of the logarithm is used, *I* is expressed in bits of information. Furthermore, it can be shown that *I* is necessarily equal or larger than zero (Theil, 1972, at pp. 59 f.). This constraint of a non-negative *aggregated* value for *I* allows for local entropy-changes as contributions which are negative. For evolution-theoretical reasons, in a complex network one expects local structures to contribute to the redundancy. However, these negative contributions have to be normalized with reference to a relevant system that produces probabilistic entropy while developing (Leydesdorff, 1995).

The expected information is contained in a message that is received by a system *a posteriori*. In other words, the evolutionary analysis changes the time horizon to the operation of the system of reference in the present as building *a posteriori* upon the historical time-series. What does the past mean for the present? Was the system perhaps redefined by path-dependent transitions (Frenken & Leydesdorff, 2000)?

For example, what was considered 'biotechnology' in 1980 is no longer necessarily defined the same way in later years (Nederhof, 1988). For the prospective policy analysis, however, the current understanding is more relevant than a previous understanding. In other words, the historical axis is inverted when using an evolutionary perspective: the system of reference is *ex post*, whereas the historical analysis tends to fix the framework *ex ante* (Narin, 1976). Data becoming available in each year provide a potential update value for historically evolving expectations.

I shall focus below on change between the two latest available years, that is, change contained in 1999 data with respect to 1998 data. Comparisons with earlier years remain possible, in principle, and often desirable for substantive reasons—that is, for a historical understanding—but these extensions do not add fundamentally to the methodological problems under study.

## 4. Materials

The *Journal Citation Reports* of the *Science Citation Index* list the aggregated citation data of 5,550 journals in 1999 versus 5,467 in 1998. This data was reorganized in order to fit legacy software developed for the analysis of similar data in the 1980s (Leydesdorff & Cozzens, 1993). In general, citation data can be analyzed from the 'cited' and from the 'citing' side. The cell values of the grand matrix can be considered as the mutual information between these two dimensions of the matrix.

Note that the *Science Citation Index* is generated by processing the publications from the 'citing' side. Literature from the current year is scanned for references to literature in the archives. Then, the matrix is transposed in order to consider also the 'cited' dimension (Wouters, 1999). This operation in itself adds no data to the database. 'Cited', however, are also a number of journals other than those processed by the Institute of Scientific Information (ISI), the producer of these databases. For example, in 1999 194,786 items were cited by the citing documents in a total of 20,050,851 citations. The total number of cited items within the domain of the ISI journals was only 15,898,944 (that is, 79.3%).

I limit the analysis here below to the journals which were processed by the ISI both on the citing and the cited side. This reduces the number of cited references in the distribution considerably, but not in proportion to the above figures. Of the 1,371,216 unique references contained in the 1999 database, 600,171 point to source materials which were not processed by the ISI from the citing side. I will work with the remaining 771,045 citation relations (56.2%) which contain a total of 14,264,510 citations (that is, 89.7% of the total cited). (The other 10.3% are single citation relations which are subsumed by the ISI under the category 'All others'.)

|  | *1999* | *1998* | *1999/1998* |
|---|---:|---:|---:|
| number of source journals processed | 5500 | 5467 | 1.006 |
| number of items referenced | 194,786 | 187,830 | 1.037 |
| number of citation relations | 1,371,216 | 1,313,012 | 1.044 |
| total citations 'citing' | 20,050,851 | 19,227,581 | 1.043 |
| citation relations to source journal | 771,045 | 732,842 | 1.052 |
| source journal not processed 'citing' | 21 | 23 | 0.913 |
| total 'cited' | 15,898,944 | 14,920,338 | 1.065 |
| total covered by our analysis[1] | 14,264,510 | 13,771,315 | 1.036 |

**Table 1**
*Comparison of the data in various relevant dimensions for 1999 and 1998, respectively. (The 20+ source journals which were not processed on the citing side will be included into the analysis when the focus is on the 'cited' dimension.)*

Table 1 summarizes the data for 1998 and 1999. Additionally, the ISI listed 186 changes of journal names in 1999, including 6 splittings and 60 mergers between journals. The other 120 records point to name changes. I controlled for these name changes, but not for the mergers and splittings. Taking this list into account, I was able to match 5331 journals of the 5467 journals listed in 1998 (97.5%). In summary, 136 journals (2.5%) were dropped from the database between 1998 and 1999, while 83 journals (1.5%) were added.

All journals are attributed by the ISI to one or more categories in a disciplinary classification scheme. One hundred sixty such categories were specified in 1999. The 5550 journals are attributed to 8752 of these classifications, that is, 1.6 per journal on average. One can consider a set of journals with a unique classification category as a macro-journal (Cozzens & Leydesdorff, 1993). Since probabilistic entropy measures can be aggregated (given the $\Sigma$ in the Shannon-formula; cf. Theil, 1972), I will be able to specify values for these macro-journals on the basis of probabilistic entropy measures for individual journals.

**5. Methods**

If one conceptualizes the aggregated journal-journal citations as a huge matrix of 5550 journals cited versus (the same) 5550 journals citing, this matrix contains $5550^2 = 30,802,500$ cells. Whereas we have 771,045 unique citation relations (in 1999), only 2.5% of these cells contain a non-missing value. Almost all (97.5 %) of the cells are empty. Since citation patterns are highly similar within specialties, this emptiness means that the multi-dimensional space corresponding to the matrix representation can be considered as virtually empty. For evolutionary reasons, one can also expect this complex system to be nearly decomposable (Simon, 1969).

---

[1] Single citation relations are compiled by the ISI under the heading 'all others' and not included in our analysis.

From a purely statistical perspective, one implication of this relative emptiness is that the 1999 matrix is very similar to that of 1998, notably since both sets are overwhelmingly empty. Furthermore, we can only make comparisons among journals which were present in both years. In other words, the overall pattern can be expected to be rather similar when analyzed at the aggregated level. One expects occasional change or, in other words, change can be considered as an exception. Can it also be used as an indicator of newness, obsolescence, etc.?

As noted, the change of a distribution can be measured in bits of information using *I* as defined above. *I* is a non-parametric and aggregative measure. The measurement is normalized in terms of the *a posteriori* event, that is, the information is evaluated from a hindsight perspective. The multivariate extension of the dynamic entropy measure to $I = \Sigma\ q_{ijk..}\ ^2\log\ (q_{ijk..}\ /\ p_{ijk..})$ is straightforward.

Once the information is brought under the control of a database manager, several options for developing indicators using *I* can be distinguished analytically. I shall first compute the contribution of each journal to the overall change of the aggregated journal-journal citations in both the cited and the citing dimension. Thus, we will be able to specify the change in the distribution of total citations in either dimension between the two years, and this change can then be decomposed in terms of the contributions of individual journals to it ($\Delta\ I$). However, the overall change in the distribution of citation patterns among journals does not yet inform us about the change of the citation patterns of each individual journal as a (one-lower-level) vector of this matrix.

In other words, this first measure provides us with statistics which are normalized in terms of the database. They can be compared with the impact factor, but they are a measure of the dynamics whereas the impact factor is measured for each year separately. Journals can be compared directly also in terms of this indicator, since the values are normalized. A contribution to this overall change in the distribution can be expressed as a $\Delta\ I$ for each individual journal, both in the cited and in the citing dimension. However, if we wish to use the journal citation pattern as indicators of cognitive change, we need to know which journals are cited differently from the year before by each journal separately. In this case, the analysis should be performed at the level of the 771,045 cell values within the matrix in comparison to the 732,842 values available in 1998.

Furthermore, one can distinguish between the probabilistic entropy generated at the level of each vector and the probabilistic entropy generated at the level of the matrix by specific citation interactions aggregated for each journal. The latter case implies a normalization. Thereafter, journals can again be compared and aggregated. When comparing vectors, however, the values for different journals cannot be aggregated. The size of the journal may affect the analysis. We will pursue both types of analysis here below, yet with a focus on the 'cited' side. (In a later study, I hope to return to the specifics of the 'citing' dimensions or the combination of cited/citing as indicators of novelty.)

In order to keep the problem computationally tractable, let me first rewrite the formula for *I* in the following way:

$$I\quad =\quad \Sigma\ q_i\ ^2\log\ (q_i\ /\ p_i)$$

By writing $\Sigma\ q_i$ and $\Sigma\ p_i$ as relative frequency distributions:

$$\Sigma\, q_i = \Sigma\, f_q/n_q \quad \text{and} \quad \Sigma\, p_i = \Sigma\, f_p/n_p \,, \text{respectively}$$

$$I = \Sigma\, f_q/n_q \; {}^2\!\log \{(f_q/n_q) / (f_p/n_p)\}$$

$$= \Sigma\, f_q/n_q \; \{{}^2\!\log (n_p/n_q) + {}^2\!\log (f_q/f_p)\}$$

$$= ({}^2\!\log n_p - {}^2\!\log n_q) + (1/n_q) \{\Sigma\, f_q \, {}^2\!\log (f_q/f_p)\}$$

The right hand-term enables us to operate directly on the comparable cell values as relative frequencies. The normalization into relative frequency distributions can then be performed after this addition is completed, since the summation for $n_q$ and $n_p$ can be computed in the same pass of the computer program as the computation for the right-hand term which represents the dynamics. Note that $n_q$ and $n_p$ are different for each vector, but at the level of the complete database or matrix $n_q$ and $n_p$ are constants. In the latter case one can therefore use $\Sigma\, f_q \log (f_q/f_p)$ directly as an indicator of change.

In addition to the total number of citations in each year ($n_p$ and $n_q$), the number of journals involved in the citation process of each journal under study provides us with a third parameter for the normalization. This journal-specific citation window limits the width of the channel that can be used for producing probabilistic entropy. I shall indicate this number below with N. We will explore normalization using this number of journals (N), but also the ${}^2\!\log(N)$, since the log-value normalizes with reference to the maximum information capacity of the communication channel under study.

Let me provide an example in order to explain in greater detail what I will do. Assume that journal A is cited in the year 1998 by Journals B, C, D and E. In 1999, Journal A is cited by Journals C, D, E, and F. The analysis focuses on the number of citations by Journals B, C, and D in this case, since these citations can be compared as relative frequency distributions. On the one hand, the inclusion of Journal F would lead to a division by zero (in the *a priori* cell) and therefore an infinite information value: the citation of Journal A by Journal F in 1999 can be considered as unpredictable in terms of the 1998 expectation. Only on its second occurrence can a citation be evaluated with reference to structural change. On the other hand, the disappearance of Journal B from the citation pattern of Journal A leads to a zero in the denominator and therefore to a term which is equal to zero by definition ($0 \log 0 \equiv 0$). In other words, this disappearance does not add information to our expectation about what will happen next and, therefore, it does not add to the value of the dynamic indicator in the present (1999).

In a previous study, I experimented extensively with a focus on new journals, but the results were not satisfying (Leydesdorff, 1994). New journals are added to the database both because existing fields can expand and because of new developments (cf. Garfield, 1990). The quest is here for an indicator which picks up the signal of structural change in the (citation) distribution pattern among journals, but not restricted to the inclusion of new journals in the database.

## 6. Results

### 6.1 'Cited' and 'citing' at the level of the journal-journal citation network

Since 5331 of the 5550 journals included in 1999 could be matched with journals in 1998, one would expect only these 5331 journals to contribute to the change in the overall citation pattern on the cited side. (As noted, another 20+ journals were not included as 'citing'.) The *I* generated among the cited distributions between these two years was 24.324 millibits, while the total $I_{citing}$ = 87.926 millibits, or more than three times as much. This result is consistent with the theoretical notion that the 'cited' side represents the archive of the journals, while 'citing' can be considered as the running operator generating the variation. Elsewhere I have discussed this in terms of citing as an action parameter, while the cited journals can be considered as providing an indicator of structural change (Leydesdorff, 1995).

Although the total of probabilistic entropy (*I*) produced by change is necessarily larger than zero—in correspondence with the second law of thermodynamics (cf. Theil, 1972, at pp. 59f.)—the relative contribution of each term to the summation can be positive or negative indicating an increasing or decreasing contribution to the dividedness of the distribution at the set level. The number of journals which contribute *positively* to the overall change between these two years is 2375 on the citing side, while it is 3238 on the cited side. Consequently, the journals which make a contribution to the process of change by citing are more specific than the journals which contribute to change on the cited side. In other words, the citing side *generates* structure by selective citation (Fujigaki, 1998) while the archival structure dissipates more slowly over time.

Table 2 exhibits the top twenty journals in terms of their contribution to change in both the cited and citing dimensions. These journals are, in other words, sorted in terms of the $\Delta I$ to the change in the distribution of the total citations on either side.

| *cited* | *citing* |
|---|---|
| APPL PHYS LETT | SEMIN THROMB HEMOST |
| ASTROPHYS J | ADV GENET |
| J APPL PHYS | SEMIN NEUROL |
| J PHYS CHEM B | HOSP MED |
| J PHYS CHEM A | ADV MAR BIOL |
| PHYS REV LETT | IEEE T APPL SUPERCON |
| MOL CELL | PARASITE IMMUNOL |
| J NEUROSCI | PHYS CHEM CHEM PHYS |
| APPL OPTICS | J TOXICOL ENV HEAL A |
| MON NOT R ASTRON SOC | OSTEOPOROSIS INT |
| NAT MED | CHINESE CHEM LETT |
| EUR PHYS J C | J PHARM PHARMACOL |
| NAT NEUROSCI | JMRI-J MAGN RESON IM |
| PHYS REV B | SPORTS MED ARTHROSC |
| ASTRON ASTROPHYS | SEMICONDUCT SEMIMET |
| PHYS REV E | MED ONCOL |
| CURR BIOL | ADV CHEM PHYS |
| PHYS REV D | SKULL BASE SURG |
| AM J RESP CRIT CARE | HYDROBIOLIGIA |
| J ELECTROCHEM SOC | J MATER SCI LETT |

**Table 2**
*Twenty journals contributing most to the change of the overall citation pattern between 1998 and 1999, both in terms of 'being cited' and 'citing.'*

Let us now consider in greater detail whether these journals can serve as indicators of structural change. As noted, this will be done by focusing on the 'cited'-side, since this dimension represents the archival structure, while 'citing' represents the potentially more volatile running parameter.

Using the first journal on the 'cited' side of the list (that is, *Applied Physics Letters*) as a seed journal, structural change is found indeed between 1998 and 1999. In 1998, this journal's loading pattern for 'being cited' is interfactorially complex on a second factor indicating a cluster of 'applied physics journals' (with the *Journal of Applied Physics* as the leading journal), and a fifth factor indicating journals which use physics in an applicational context, such as the *Journal of Crystal Growth*, the *Journal of Electronic Materials, Materials Sciences and Engineering B*, and the *Japanese Journal of Applied Physics*.

In 1999, the citation patterns of these two groups of journals were merged into a single cluster, whereas the factor laodings of these same journals were sometimes negatively correlated in 1998. In 1998, for example, the leading journal on Factor II (*Journal of Applied Physics*) loaded also negatively on the fifth factor (e.g., the *Journal of Crystal Growth*). In 1999, this interfactorial complexity around *Applied Physics Letters* is completely resolved when one takes this journal as a seed journal. This result suggests an increase in the codification related notably to the *Letters* exchange between these two fields. The two areas melted together as frames of reference in an otherwise stable environment of solid state physics, optics, surface and vacuum sciences, etc.

The analysis of the second journal of Table 2 (that is, *Astrophysics Journal*) provides us with an very similar picture for 1998 and 1999. The respective MDS-projections are provided in Figure 1. (The solution for 1998 is inverted in order to show the similarity between the two pictures.)

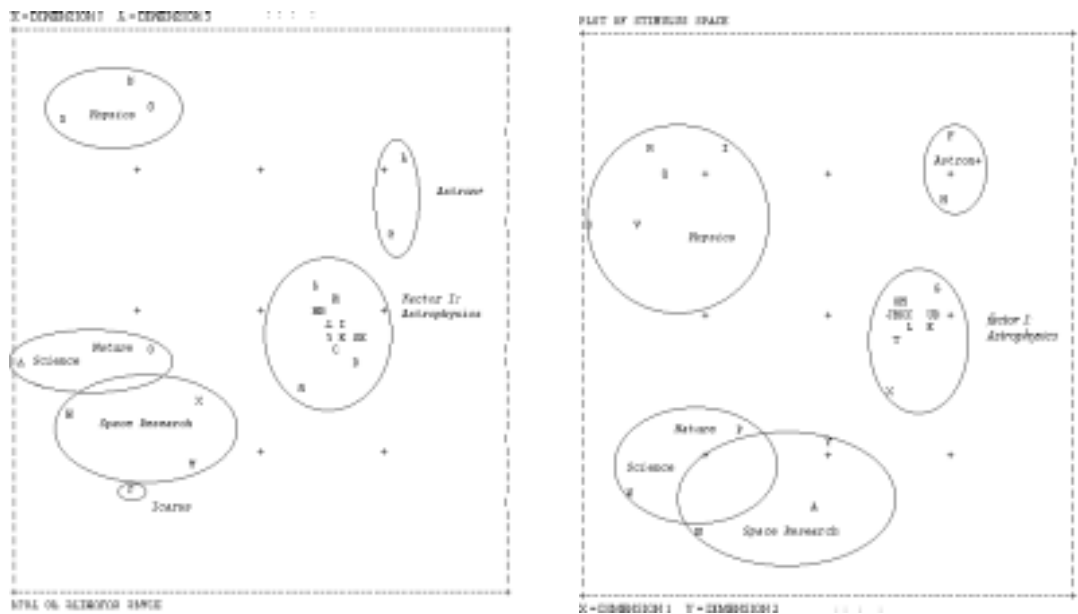

**Figure 1**
*Comparison of Multi-dimensional scaling solutions in two dimensions using Astrophysics Journal as a seed journal for the citation environment in 1998 (to the left) and 1999 (to the right), respectively. (The solution for 1998 was inverted in order to show the similarity between the two pictures.)*

In this case change seems to be observable only at the level of the database, but not in the specific citation pattern of journals. *Astrophysics Journal* has been disproportionally more cited over the file and also in its direct environment, but further analysis indicates no change in positions at the level of the specialty or among the relevant specialties in its environment. For a systematic indication of developments at the specialty level, one obviously has to zoom in on a one-lower level of analysis, that is, the specific citation pattern of individual journals.

*6.2     Citation patterns at the level of individual journals*

The previous analysis assessed change at the level of the database. However, we know from the literature that specialties differ considerably in terms of citation behaviour. For example, the journal mappings in Figure 1 (above) are based on the specificity of the citation patterns among journals belonging to astrophysics and its environment. However, the problem is that we have no obvious access to groupings over the file other than by an *ex ante* classification scheme. In terms of data, however, we have only journal-journal citation data which can be analyzed either at the level of the file or at the level of changes in the citation behaviour of individual journals. From this empirical perspective, attributions to groupings remain inferences about limited sets.

Can one use the citation behaviour of *individual* journals as indicators of change at the specialty level? Operationally, this question requires the assessment of change at the level of individual rows and/or columns of the matrices. In other words, we lower the level of aggregation by one step. Although citation practices can be expected to be much more uniform at the level of specialties, we remain yet uncertain of the normalization because the specific environments cannot be clearly delineated. All groupings remain inferential and provisional.

Normalization in terms of the total number of citations of journals ($n_p$ and $n_q$) is already implied in the formula for probabilistic entropy, since the measure operates on probabilities, that is, on *relative* frequencies. The number of journals used for citation (N) can be conceptualized as the width of the communication channel or its maximum entropy. If all citation rates to or from individual journals by the citing (or cited) journal were equal, maximum entropy would have been reached. This maximum entropy is equal to $^2\log(N)$, where N stands for the number of cited or citing journals as categories. As noted, I will explore the use of normalization both in terms of N and in terms of the logarithm of N.

For example, general science journals (like *Science*, *Nature*, and *PNAS*) cite (and are cited by) a large number of journals from a range of disciplines. While all these disciplines are developing, and the foci of attention within these journals are also changing, one would expect a lot of probabilistic entropy to be generated between two subsequent years because of differences in the news value produced by a variety of specialties. In other words, the probabilistic entropy generated in the citation patterns of these general science journals can be expected to have various origins and the specificity is diluted by its dissolution into the larger pool of citations. If a specialist journal with a narrow citation window is changing significantly in terms of its citation patterns, this may produce a relatively smaller amount of probabilistic entropy, but this smaller amount of information may provide us with more probabilistic entropy per citing or cited journal.

Let me first turn to the non-normalized case (see Table 3). Which journals produced most entropy in relation to their citation profile in the previous year using the files of 1999 as against 1998? One can then raise further questions about the origins of this probabilistic change by zooming in on the difference in the citation mappings between these two years.

| cited journals | sorted on probabilistic entropy production (1999\|1998) | N = |
|---|---:|---:|
| RESTOR NEUROL NEUROS | 2.019 | 26 |
| T I MIN METALL B | 1.819 | 8 |
| APPL SUPERCOND | 1.631 | 14 |
| J NON-EQUIL THERMODY | 1.471 | 12 |
| CRYOGENICS | 1.385 | 42 |
| CIM BULL | 1.379 | 17 |
| J APPL STAT | 1.357 | 12 |
| ROBOT AUTON SYST | 1.275 | 5 |
| T INDIAN I METALS | 1.224 | 10 |
| CHILD NEUROPSYCHOL | 1.200 | 2 |
| IEEE T APPL SUPERCON | 1.161 | 47 |
| INT J GEN SYST | 1.102 | 10 |
| OSTEOPOROSIS INT | 1.101 | 102 |
| FOLIA ZOOL | 1.091 | 12 |
| INT J ENVIRON POLLUT | 1.085 | 2 |
| SUPERCOND SCI TECH | 1.074 | 58 |
| CRYPTOGAMIE MYCOL | 1.032 | 4 |
| REV FR ALLERGOL | 0.979 | 10 |
| J AQUAT PLANT MANAGE | 0.963 | 5 |
| J SYN ORG CHEM JPN | 0.958 | 40 |

**Table 3**
*Journals with changing citation patterns in the cited dimension in decreasing order.*

Let us analyze the first ten journals or so in detail using a matrix at the 1% level of the total citations, and by factor analyzing and multidimensional scaling this matrix in the cited dimension. These methods are explained more fully in Leydesdorff & Cozzens (1993) and Leydesdorff *et al.* (1994).

*6.2.1 Restorative Neurology and Neuroscience*

By using this journal as the seed for the delineation of a journal environment, a new journal *cluster* can be made visible in 1999 which was not present in 1998. While *Restorative Neurology and Neuroscience* loaded on the first 'neuroscience' factor in 1998, it loads as a central tendency journal with the *Journal of Neurotrauma* on a separate (twelfth) factor in 1999. Its factor loading on the first (otherwise stable) factor 'neuroscience' decreases from 0.774 in 1998 to 0.024 in 1999, that is, this correlation approaches zero. In other words, the emerging citation pattern branches off in an orthogonal direction.

Figure 2 exhibits the resulting journal structure emerging between clusters of journals indicating 'neuroscience,' 'cell biology,' 'neurology,' 'neurophysiology,' 'neurosurgery,' 'brain injury research,' 'vision research,' and 'behavioural brain research.' Since the seed journal (*Restorative Neurology and Neuroscience*) generates a citation environment of 124 journals in 1998 and 98 journals in 1999 using our default for the threshold of 1% of the totals cited and citing, the picture is based on raising this threshold to 2%. With this threshold, 49 journals are included in the citation environment.

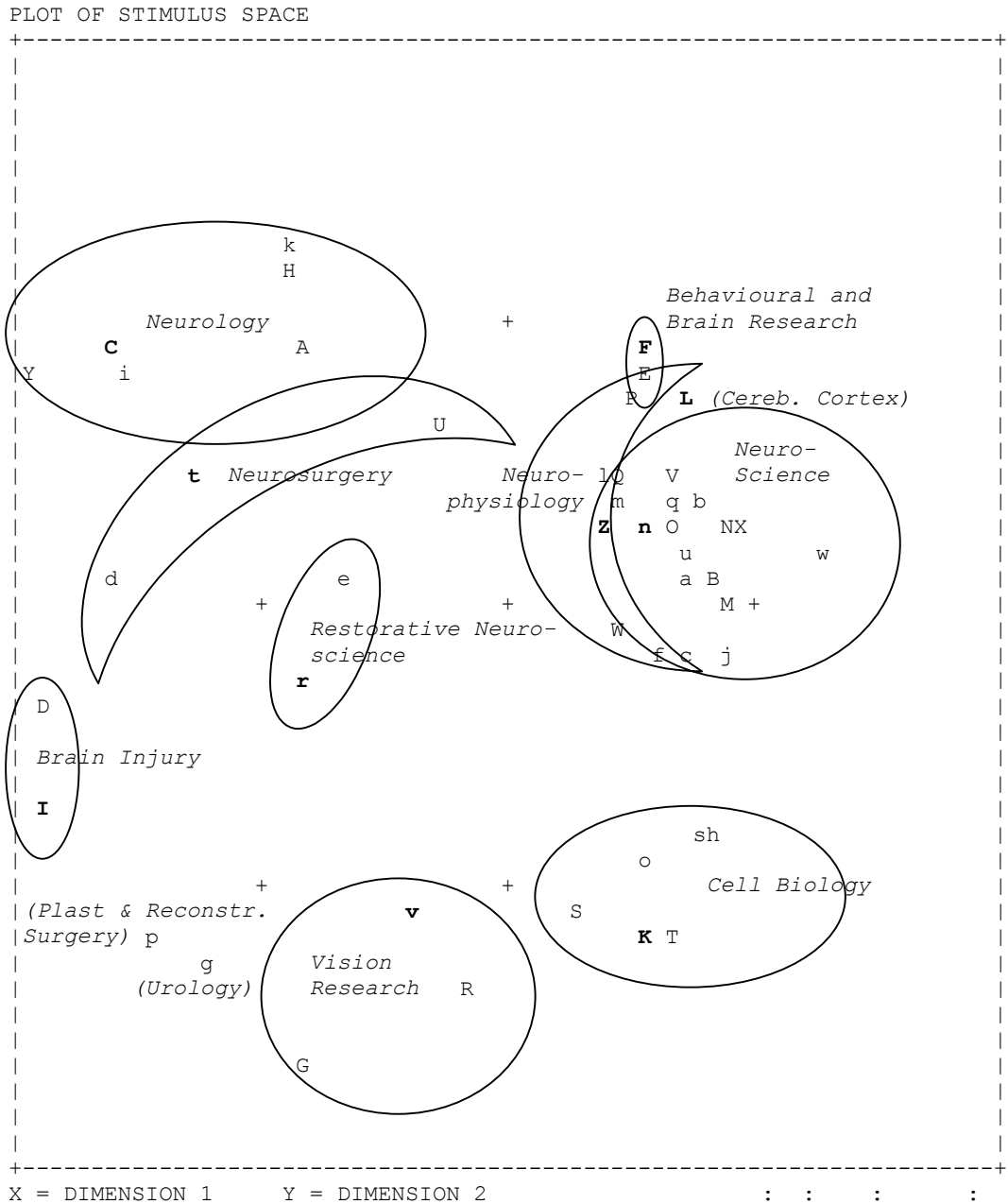

**Figure 2**
Factor-analysis and MD-SCAL for *Restorative Neurology and Neuroscience*, cited patterns
(1999; threshold = 2.00%)[2]

---

[2] Journals with highest factor loadings are given in boldface both in the figure and in the legend. Note that some journals are percieved as 'neuroscience' by the citation environment, while they may identify themselves as 'neurology' both in their names and in their citing patterns.

| journal name | factor | journal name | factor |
|---|---|---|---|
| A. ANN NEUROL | III | a. J NEUROSCI | I |
| B. ANNU REV NEUROSCI | I | b. J NEUROSCI METH | I |
| **C. ARCH NEUROL-CHICAGO** | **III** | c. J NEUROSCI RES | I |
| D. ARCH PHYS MED REHAB | X | d. J NEUROSURG | V |
| E. BEHAV BRAIN RES | VII | e. J NEUROTRAUM | XI |
| **F. BEHAV NEUROSCI** | **VII** | f. J PHYSIOL-LONDON | IV |
| G. BIOORG MED CHEM LETT | XII | g. J UROLOGY | |
| H. BRAIN | III | h. NATURE | II |
| **I. BRAIN INJURY** | **X** | i. NEUROLOGY | III |
| J. BRAIN RES | I | j. NEURON | I |
| **K. CELL** | **II** | k. NEUROPSYCHOLOGIA | III |
| **L. CEREB CORTEX** | **VIII/IX** | l. NEUROREPORT | I |
| M. CURR OPIN NEUROBIOL | I | m. NEUROSCI LETT | I |
| N. DEV BRAIN RES | I | **n. NEUROSCIENCE** | **I** |
| O. EUR J NEUROSCI | I | o. P NATL ACAD SCI USA | II |
| P. EXP BRAIN RES | IV | p. PLAST RECONSTR SURG | |
| Q. EXP NEUROL | I | q. PROG NEUROBIOL | I |
| R. INVEST OPHTH VIS SCI | VI | **r. RESTOR NEUROL NEUROS** | **XI** |
| S. J BIOL CHEM | II | s. SCIENCE | II |
| T. J CELL BIOL | II | **t. STROKE** | **V** |
| U. J CEREBR BLOOD F MET | V | u. TRENDS NEUROSCI | I |
| V. J COMP NEUROL | I | **v. VISION RES** | **VI** |
| W. J NEUROCHEM | I | w. VISUAL NEUROSCI | VI |
| X. J NEUROCYTOL | I | | |
| Y. J NEUROL NEUROSUR PS | III | | |
| **Z. J NEUROPHYSIOL** | **IV** | | |

Can we also find a means to check from another angle whether this is indeed an emerging development? By using the *Journal of Neurotrauma*, that is, the other journal loading on the emerging cluster, one is able to generate another perspective on this same field in 1998 and 1999. In 1998, the journal *Restorative Neurology and Neuroscience* was not yet present in the citation environment of the *Journal of Neurotrauma*, but it is in 1999. However, the two journals from this perspective load on different factors, while the *Journal of Neurotrauma* also exhibits interfactorial complexity in this year.[3] In summary, the new development is not sufficiently codified to be noted when the perspective of the *Journal of Neurotrauma* is used, while the emergence of a new journal cluster is visible from the journal which was clearly indicated by our information measure.

### 6.2.2   *Transactions of the Institution of Mining and Metallurgy, Section B*

In 1999, this journal (*T I Min Metall B*) becomes part of a cluster that consists also of *Mineralium Deposita*, *Economic Geology and the Bulletin of the Society of Economic Geologists*, and *Ore Geology Reviews*. Other clusters in the environment indicate 'geology,' 'mineralogy,' 'chemical geology,' and 'sedimentology'. In 1998, this same journal functions as an isolate in a context which is dominated by a geophysics factor which is absent in the 1999 solution, whereas the other clusters are present in both years. Thus, it seems in this case that we are witnessing the inclusion of an existing journal which further influenced the journal structure by splitting it into a more applied and a more basic part.

---

[3] Van den Besselaar & Heimeriks (2001) argue that interfactorial complexity can be considered as an indicator of interdisciplinarity.

Let us analyze this development in more detail and examine whether using the journal with the largest loading on the 1999-cluster (*Mineralium Deposita*) as a seed journal, also indicates structural change? In both 1998 and 1999, *Mineralium Deposita* is a 'central tendency journal' in the cited dimension. The environment of this cluster is relatively stable, but its internal composition changes. In 1998, the cluster was further composed of *Ore Geol Rev*, *Econ Geol Bull Soc*, the *Canadian Bulletin of Mining and Metallurgy (CIM Bull)*, and the *Australian Journals of Earth Sciences*, while in 1999 *T I Min Metall B*, *Econ Geol Bull Soc*, and *Ore Geol Rev* formed this cluster. Thus, two journals with a national identity in the title (*CIM Bull* as a Canadian journal and the *Australian J Earth Sciences*) no longer load on this factor in 1999.

Note that *CIM Bull* (that is, the *Canadian Bulletin of Mining and Metallurgy*) is in the sixth position on the list generated by our indicator (Table 3). In 1999 this journal exhibits the highest factor loading in its citation environment on an eigenvector which further covers *Exploration and Mining Geology* and *T I Min Metall A*. In 1998, however, *CIM Bull* did not play this role in the codification structure; it was then part of a third cluster which can be identified as 'mineralogical engineering' (the *Engineering and Mining Journal* and *Minerals Engineering*), while in this citation environment *Mineralium Deposita* exhibits the highest factor loading on the first eigenvector again together with *Ore Geol Rev* and *Econ Geol Bull.*

In summary, we are witnessing a reorganization of the codification structure of this field in which the repositioning of *CIM Bulletin* may have played a central role. In the 1999 environment (as against the 1998 environment) of this journal, another journal becomes visible which is also on our list in Table 3, notably *T Indian I Metals*. This journal is on the ninth position using our indicator. It loads in this environment (of *CIM Bulletin*) on a fourth factor together (but behind) *JOM-Journal of the Minerals, Metals and Materials Society* and *Hydrometallurgy*.

The *Transactions of the Indian Institute of Metals* is one of those journals which entertain a citation relation with a large number of journals. At the 1% percent threshold 65 journals were drawn into its citation environment in 1998, and 112 journals in 1999. Perhaps, the journal may function as a window on this environment from an Indian perspective. In this citation environment *CIM Bull* loads as a second variable on a factor with *Hydrometallurgy* and *ISIJ International*.[4] Major factors in the environment can be designated as 'materials science and technology,' 'corrosion,' 'electrochemistry,' and 'applied physics.'

In 1998, 'corrosion' was not yet a factor in this journal environment (at the 2% level). 'Mineral processing' had this place in the earlier year. 'Ceramics' was also more prominently present. Perhaps we can observe here a representation of the diffusion of 'new materials' in relation to more traditional 'raw' materials and the impact of this techno-science on the relevant journal environments. The issue of corrosion of materials seems to play a role as well. In summary, this development is traced by our indicator as very important in terms of changing citation patterns among various fields of science. Three journals relevant to this field were flagged by our indicator. An in-depth analysis would require back-tracking the changes in these citation patterns to earlier years.

*6.2.3 Applied Superconductivity*

The third journal indicated in Table 3 is *Applied Superconductivity*. Additionally, the eleventh journal on our list is *IEEE Transactions on Applied Superconductivity*, and the fifth one is

---

[4] The abbreviation *ISIJ* stands originally for the 'Iron and Steel Institute of Japan.'

*Cryogenics*. These three journals participate in each other's citation environments and they indicate a pattern of change between 1998 and 1999. The probabilistic entropies for these three journals together generate 4.176 bits of information, that is, more than twice as much as the top journal on the indicator (*Restorative Neurology and Neuroscience*) which we discussed above (in section 6.2.1). Note that an indicator at the journal level may indicate a change at the cluster level which can still be of variable size.

The journal *Applied Superconductivity* loads highest on the third factor in its citation environment both in 1998 and in 1999. It can be considered as a central tendency journal in both these years, and the cluster itself also exhibits stability. This third factor explains 12.3% of the variance in the matrix in 1998 and 12.9% in 1999. In other words, the factor solutions are virtually identical, with the only difference in the relative change in the position of *Cryogenics*.

| 1998 | factor loading | 1999 | factor loading |
|---|---|---|---|
| *Appl Supercond* | 0.96283 | *Appl Supercond* | 0.96614 |
| *IEEE T Appl Supercond* | 0.92651 | *IEEE T Appl Supercond* | 0.90872 |
| *Supercond Sci Tech* | 0.91812 | *Supercond Sci Tech* | 0.90654 |
| *Physica C* | 0.75933 | *Cryogenics* | 0.78660 |
| *Cryogenics* | 0.53051 | *Physica C* | 0.60360 |

**Table 4**
*Factor Three ('superconductivity') with major factor loadings in 1998 and 1999, respectively.*

The shift therefore is rather to be found in the organization of the environment of this cluster and its position in the database. In 1998, the first cluster in the citation environment of *Applied Superconductivity* can be designated as solid state physics and condensed matter physics (among which is the journal *Physical Review B*). The same cluster can be found in 1999, but the internal order in the cluster is now less indicative of condensed matter physics as a specialty and more of physics as a discipline. Thus, the cluster has further integrated the 'superconductivity' group into the core of the discipline.

In the second factor, a similar change can be noted, but also in this case change is relative: the 'material science' journals which led the cluster in 1998 are secondary in 1999. The 'applied physics' journals take the lead of this cluster in the later year. In minor factors there is some change, but this can also be an artifact of our procedures (e.g., cut-off levels).

In summary, using *Applied Superconductivity* as the seed journal we did not find structural change at the network level, but rather a gradual change in the direction of basic physics. What is happening? Let us first turn to *Cryogenics* which—as noted—is also the fifth journal on our indicator. Used as a seed journal, this journal in 1998 loaded with 0.49531 on a second factor led by *IEEE Transactions of Applied Superconductivity*. However, the journal *Applied Superconductivity* was not part of its citation environment in this year. The first factor in 1998 can be designated as 'material science and engineering,' while the third factor is recognizable as 'solid state physics'.

In 1999, *Cryogenics* loads with 0.77449 on the first ('superconductivity') factor with *IEEE Transactions of Applied Superconductivity* (0.90170) and *Superconductor Science and Techology* (0.82875). The second factor is led by the *Journal of Low Temperature Physics* like the third factor in the previous case, while the third factor is now composed of the *J of Heat Transfer-Transactions of the ASME* and the *International Journal of Heat and Mass Transfer*. A fourth factor is 'applied physics'. In other words, this journal has made an upward movement in the journal hierarchy. However, *Applied Superconductivity* itself is not visible in this citation environment in either 1998 or 1999 (at the 1% level). The distributions of these citation environments are very skewed: thus, the journal exhibits a high specificity more typical of journals in the physics domain.

Finally, let us take *IEEE Transactions of Applied Superconductivity* as a seed journal. In both years under study this journal loads on a second factor with *Applied Superconductivity* in the top position in 1999, while it leads the same factor in 1998. *Cryogenics*, however, is part of this cluster in 1999, while it was not in 1998, when it loaded from this perspective as an isolate on a lower-level factor.

The order of the factors in the citation environment of *IEEE Transactions of Applied Superconductivity* has also changed over these two years. In 1998, the factor 'superconductivity' was sandwiched between a Factor I consisting of journals in applied physics and material science and a Factor III led by the *Journal of Superconductivity*, but which can be recognized as a physics cluster containing also *Physical Review B* and *Physica C*. In 1999, the relevant citation environments are *first* physics itself, secondly applied physics (the third factor), and then a factor consisting of *Science* and *Nature*. A fifth factor has a focus on 'magnetism', while a final factor refers to instrument journals. Thus, this field has become embedded in a more purely physics-oriented environment.

In other words, we are witnessing here the reorganization of the 'superconductivity' field from a specialty on the applied side of physics moving up the discipline hierarchy to become established in the core of the discipline itself. This is, for example, evident from the emergence of journal clusters in its citation environment with a more applied character than 'superconductivity' itself. The change of position of *Cryogenics* as part of this cited structure seems to have been crucial to the transition. *Cryogenics* became visible as a codified component of superconductivity research in the later year, while this journal was the main communication channel with other applicational fields before.

Let me note that this change cannot easily be retrieved by using a comparative static analysis. The change in the relationships are entailed in a restructuring and reorientation within the relevant substructures of physics.

### 6.2.4  *The Journal of Non-equilibrium Thermodynamics, etc.*

The fourth journal on the list is the *Journal of Non-equilibrium Thermodynamics*. The 'cited' pattern of this journal exhibits change, but this citation pattern is weakly codified. When used as an entrance journal in 1998, the one-percent threshold draws 61 journals into the analysis. Limiting the threshold to 2% provides us with a citation environment of 37 journals that distribute over 13 factors with eigenvalues larger than one.[5] The journal loads highest (0.68886) on Factor XI behind *Continuum Mechanics and Thermodynamics* (0.73992). In 1999, this latter journal is no longer present in the citation environment of the *J Non-Equil Thermodyn* even if one lowers the threshold to one percent. The journal is then an isolate in a citation environment of 46 journals, loading only on Factor XII. In summary, we observe here a journal which lacks codification in its citation pattern and therefore generates noise in the database.

---

[5] An eigenvector with a value larger than one explains more than an average variable. The value of one is therefore often used as a cut-off point (e.g., in SPSS).

Let me limit the discussion about the other journals on the list to the main points. The *Journal of Applied Statistics* which is firmly embedded in a cluster of ecology journals, takes the lead for this journal cluster in the cited dimension in 1999, while it was in the last position in 1998. The journal *Robotics and Autonomous Systems* is part of a cluster which seems to emancipate itself from robotics and pattern recognition research. The latter journals are no longer in its citation environment in the later year, and its relation with artificial intelligence has become more pronounced than with robotics.

Finally, *Child Neuropsychology* provides us with a case where a factor has disappeared. While leading as a 'central tendency journal' a marginal cluster in 1998 which was otherwise composed of the *International Journal of Neuroscience* and *Developmental Neuropsychology*, this journal became part of the citation cluster of clinical neuropsychology (Factor III) in 1999. *Developmental Neuropsychology* loads in this later year on the first factor which is focused on the cognitive side of neuropsychology. Note that only two journals cited *Child Neuropsychology* both in 1998 and in 1999. Thus, this journal has been completely replaced in the codification structure of the *Science Citation Index*.

*6.3    Normalization in terms of the bandwidth of the channel*

In several of the cases discussed above we signaled that the initial seed journal had a less codified position then the cluster to which it belonged. Change is often introduced from interaction at the margins of otherwise more stabilized clusters. Such journals may exhibit a more dispersed citation window and therefore relate to more journals. In the factor analysis, this 'interdisciplinarity' can lead to interfactorial complexity (Leydesdorff & Cozzens, 1993; Van den Besselaar & Heimeriks, 2001). It seems appropriate to check whether such an effect plays a role here by controlling for the number of journals involved in the citation relation.

| cited journal | sorted on $I / {}^2log(N)$ | sorted on $I / N$ | $N =$ |
|---|---|---|---|
| CHILD NEUROPSYCHOL | 1.200 | 0.600 | 2 |
| INT J ENVIRON POLLUT | 1.085 | 0.542 | 2 |
| QUAL QUANT | 0.748 | 0.374 | 2 |
| STRATIGR GEOL CORREL | 0.695 | 0.347 | 2 |
| MED PROBL PERFORM AR | 0.669 | 0.334 | 2 |
| MEAS CONTROL | 0.643 | 0.322 | 2 |
| T I MIN METALL B | 0.606 | 0.227 | 8 |
| ROBOT AUTON SYST | 0.549 | 0.255 | 5 |
| ITE J | 0.525 | 0.277 | 3 |
| EPRI J | 0.524 | 0.262 | 2 |
| CRYPTOGAMIE MYCOL | 0.516 | 0.258 | 4 |
| ACTA BIOL CRACOV BOT | 0.510 | 0.269 | 3 |
| ADV COMPOS MATER | 0.500 | 0.264 | 3 |
| BRENNST-WARME-KRAFT | 0.492 | 0.260 | 3 |
| J CLIN NEUROSCI | 0.488 | 0.258 | 3 |
| ANTHROZOOS | 0.477 | 0.252 | 3 |
| RESTOR NEUROL NEUROS | 0.430 | 0.078 | 26 |
| APPL SUPERCOND | 0.428 | 0.116 | 14 |
| J AQUAT PLANT MANAGE | 0.415 | 0.193 | 5 |
| EVOL HUM BEHAV | 0.412 | 0.206 | 2 |

**Table 5**
*Cited journals sorted on the value of $I / {}^2log(N)$ for the Journal Citation Report 1999 as against 1998.*

Table 5 exhibits the top list of journals cited when the indicator is normalized in terms of the $^2\log(N)$ or N, respectively. Journals that have very small citation windows tend to be sorted to the top of this list by this normalization. More detailed analysis of the top ten journals showed that the pattern described above for *Child Neuropsychology* often applies to such cases: these journals mainly register the disappearances of specific citation patterns. These disappearances can then be considered as enhanced codifications by a journal structure in which the journal under study is more strongly included than in the previous year. *This erosion of structure in the archive of science seems to be a more important aspect of overall change than the emergence of new structural components.*

*6.4    Comparison of individual citation patterns at the level of the matrix*

In addition to normalizing over the vectors, one is also able to normalize the contributions to the total entropy generated when comparing the matrix of 1999 with similar cell-values in 1998. As explained in the methods section above, the value of *I* can for that purpose be rewritten in terms of constants and a normalized summation. In Table 6, the order of the journals is shown with reference to the value of the latter term.

| cited journal (row vectors of the matrix) | $\Sigma\ f_{1999}\ ^2\log(f_{1999} / f_{1998})$ |
|---|---:|
| J BIOL CHEM | 50852 |
| NATURE | 43316 |
| SCIENCE | 40777 |
| P NATL ACAD SCI USA | 32116 |
| APPL PHYS LETT | 30805 |
| J CHEM PHYS | 29833 |
| PHYS REV B | 29297 |
| PHYS REV LETT | 25300 |
| J APPL PHYS | 19539 |
| J NEUROSCI | 19204 |
| J IMMUNOL | 18450 |
| BLOOD | 17791 |
| LANCET | 17773 |
| NEW ENGL J MED | 16295 |
| CIRCULATION | 15613 |
| CELL | 15214 |
| TETRAHEDRON-ASYMMETR | 14907 |
| CANCER RES | 14850 |
| NEUROLOGY | 12919 |
| J VIROL | 12838 |

**Table 6**
*Top twenty journals sorted in terms of the entropy production at the level of the matrix.*

The journals on this list are obviously recognizable as leading journals in the database. In other words, what one observes here is a manifestation of the so-called 'Matthew effect' in science (Merton, 1968). Leading journals profit from the erosion of fine structure in the database. The Gospel According to St. Matthew puts it this way: 'For he that hath, to him shall be given: and he that hath not, from him shall be taken even that which he hath.' (Matt. 4: 25).

In summary, selection if successful can be considered as a self-reinforcing process: the deselected cases are further deselected, and the archive of science over time becomes increasingly codified as previous variations fade away. *At the level of the full database,* this effect overshadows the possibility of perceiving new structural elements.

*6.5    Using ISI-categories as macro journals*

As can be expected, the macro-journals which can be composed on the basis of the 160 ISI categories suffer from the problem that different numbers of journals are involved. The largest category in terms of the number of journals is *Biochemistry & Molecular Biology,* with 551 journals subsumed in it, and it also generates the most probabilistic entropy between 1998 and 1999 (59.97 bits of information). Second is the category of *Neurosciences,* with 360 journals and 44.15 bits of information. The third cluster in terms of size is *Pharmacology*, but aggegated change here is slightly smaller than in the case of the fourth cluster, *Engineering, Electrical and Electronic.* In summary the rank-order correlation between the size of the cluster and the probabilistic entropy generated is very high (Spearman's $\rho = 0.97$).

| *ISI-category* | *I(cited) / N* | *number of journals N* |
|---|---|---|
| MINING & MINERAL PROCESSING | 0.232 | 31 |
| GEOLOGY | 0.197 | 56 |
| MICROSCOPY | 0.196 | 17 |
| PALEONTOLOGY | 0.195 | 51 |
| MINERALOGY | 0.179 | 43 |
| HORTICULTURE | 0.176 | 17 |
| ORNITHOLOGY | 0.170 | 24 |
| MARINE & FRESHWATER BIOLOGY | 0.169 | 126 |
| GEOGRAPHY | 0.168 | 33 |
| MATERIALS SCIENCE, CERAMICS | 0.165 | 36 |
| THERMODYNAMICS | 0.162 | 65 |
| PHYSICS, APPLIED | 0.161 | 124 |
| PERIPHERAL VASCULAR DISEASE | 0.161 | 79 |
| FORESTRY | 0.160 | 52 |
| GERIATRICS & GERONTOLOGY | 0.156 | 39 |
| ENERGY & FUELS | 0.155 | 105 |
| MYCOLOGY | 0.153 | 29 |
| GEOSCIENCES, INTERDISCIPLINARY | 0.152 | 206 |
| PHYSICS, FLUIDS & PLASMAS | 0.152 | 37 |
| ECOLOGY | 0.150 | 166 |

**Table 7**
*Twenty journal categories of ISI sorted according to the average amount of change per journal in this category (cited dimension).*

Since the probabilistic entropy of a macro-journal is based on a summation over the journals included, one can also divide by the number of journals in order to obtain a value for the (average) probabilistic entropy per journal. Table 7 lists these normalized values for the top twenty categories. *Mining and Mineral Processing* takes the first place. Actually, we discussed this relatively small cluster above (in section 6.2.2). Therefore, it is not obvious what this information adds to our understanding. The aggregation rules of the ISI follow automated attribution principles based on *ex ante* criteria that are kept stable over the years under study. The analysis at the level of individual journals is more precise and sensitive to change.

## 7. Conclusion

Using entropy statistics I have explored whether and how changes in citation patterns can be used as indicators of structural change in the database. The exploration was mainly methodological. For example, this study was restricted to change between two subsequent years. The conclusion, however, is that various structures operate as subdynamics which can be distinguished in operational terms.

For example, hierarchical codification structures headed by prominent journals like *Science* and *Nature* can be distinguished from heterarchical relations among journal clusters at the network level. The latter can be analyzed using factor analytical techniques, while the analysis of hierarchical relations requires graph analytically oriented approaches (Burt, 1982: Leydesdorff, 1995).

Different types of change could be distinguished. Change may indicate the emergence of new structural elements, but more often the production of probabilistic entropy indicates ongoing codification processes that erase previously generated structures. This erasure may result in the complete disappearance of previously discernable eigenvectors in the citation structures among journals. At the level of the network, the codification process, in terms of leading journals further accumulating high citation rates, predominates over more finely grained changes in citation patterns. However, the latter could be indicated by using the specific citation vectors *before* normalization. The normalization of these specific interaction effects at the matrix level led again to the predominance of size effects of the journals or of their citation windows, respectively, in the results of the analysis.

The journals indicated can be related to their relevant (journal-)environments in a variety of ways. We found the case of an emerging journal structure indicated by a single lead journal (*Restorative Neurology and Neuroscience*), but also a situation where the development of a new technology (advanced versus raw materials) is upsetting an existing scientific journal structure. In this case 'national' journals indicating geographical interests (of Canada, Australia, and India) also play a role in the restructuring of the scientific fields involved.

In a third case, we analyzed the scientific upgrading of a journal structure around 'superconductivity' which had witnessed an explosively expanding development on the applicational side in the earlier years (Vlachý, 1988; Leydesdorff *et al.*, 1994). This reorganization seems to be driven by a trend towards scientific codification into the more established parts of the physics discipline.

Note that these interpretations of the noted phenomena are 'thin' descriptions based on two years only. From the indicators as signals one is able to follow-up by backtracking into the relevant citation environments of these journals in previous years and provide a more detailed description of the developments indicated. It might be particularly rewarding to apply the same analysis to previous years and then to focus on the journals which indicate strong patterns of change over more than a single year. However, this would lead us into the detailed reconstruction of *historical* developments (e.g., Van den Besselaar & Leydesdorff, 1996; Leydesdorff & Van den Besselaar, 1987), while our focus here was on the development of a relatively straightforward indicator of *current* developments in the database. The historical reconstruction follows the actors using the time axis (Latour, 1987), while a policy analysis inverts the time-series by taking an evaluative perspective (Frenken & Leydesdorff, 2000).

Since all journals under study were included in the *Science Citation Index,* they have already crossed a considerable selection barrier, namely, the screening process of the ISI. Yet, a number of journals indicated lack of codification, thus representing 'noise' in the cited journal structure. These journals cannot be placed so easily. One can consider them either as a source of noise or as highly innovative. I expect more journals actively to change the aggregated position of their citations when the focus would be on the 'citing' side. On the 'citing' side, novelty and the recombination of existing citation structures can perhaps be considered as early indicators of new knowledge *claims* (Swanson, 1990; Kostoff *et al.*, 1997). In this study, however, I have focused on the use of entropy measures for the codification process of scientific knowledge in terms of 'being cited.' From an S&T policy perspective, one may expect this indicator to be relatively slow, but more robust than an indicator based on changes in the aggregated 'citing' patterns.

The dilemma of whether the changes indicated should be interpreted as noise or innovation highlights yet another important issue: the data inform us about change, but not on the quality of the changes. Innovation can only be identified in retrospect from an evaluative discourse. For example, a policy analysis may take the results of this analysis into account. Whether one wishes to focus on new cluster formation or on the further strengthening of confluences and recombinations among previously separate clusters depends on one's theory of innovative change. Methodological development and appreciative theorizing can go hand in hand.

We have seen above that change can indicate a reorganization along the axis new technology / old technology as in the case of materials sciences or, for example, in terms of applied versus basic science as in the case of the position of superconductivity journals in physics. An indicator remains just an indicator. What is being indicated when probabilistic entropy is found at certain places in the database requires a more detailed study.

My results suggest that general patterns of change cannot be expected from the indication of change in the database because the sources of change are rather heterogeneous. Processes of codification prevail at the journal level. This conclusion has also policy implications: while one may be able to sustain new developments for some time, structural change is unlikely to be achieved from a political programmatic unless the latter accords with ongoing processes of change in scientific communication (Van den Daele *et al*., 1979). Codification processes in the scientific communication structure can perhaps be considered as the longer-term selectors upon the variation generated by shorter-term S&T-policy program (Leydesdorff & Van der Schaar, 1987).

[return](#)

[return](return)